\documentclass[12pt]{iopart}
\usepackage{iopams,epsf}  
\begin{document}

\title[NJL model with infrared confinement]
{NJL model with infrared confinement}
\author{D Blaschke\dag, 
G Burau \dag, 
M K Volkov\ddag\ 
\footnote[3]{To whom correspondence should be addressed.}
and V L Yudichev\ddag}

\address{\dag\ Fachbereich Physik, Universit\"at Rostock\\
D-18051 Rostock, Germany}

\address{\ddag\ Bogoliubov Laboratory of Theoretical Physics\\
Joint Institute for Nuclear Research\\
141980 Dubna, Russian Federation}

\begin{abstract}
We consider an extended Nambu--Jona-Lasinio (NJL) model for the light meson 
sector of QCD where unphysical quark production thresholds are excluded by an 
infrared cut-off on the momentum integration within quark loop diagrams.
This chiral quark model conserves the low energy theorems.
The infrared cut-off is fixed selfconsistently by the dynamically generated 
quark mass (quark condensate).
The masses and decay widths of the $\sigma$- and $\rho$-mesons
are described in the model.
\end{abstract}


\submitted \jpg

\maketitle

\section{Introduction}
The Nambu--Jona-Lasinio (NJL) model is a convenient
semiphenomenological  quark model for the description of the low
energy meson physics \cite{eguchi,ev83,volk86,ebert86,vogl}.
Within this model the mechanism of spontaneous breaking of chiral
symmetry (SBCS) is realized in a simple and transparent way,
and the low energy theorems are fulfilled.

Unfortunately, the ordinary NJL model fails to prevent hadrons from
decaying into free quarks, which makes the realistic description of
hadron properties on their mass shell questionable. The exact solution
of this problem seems to be a very difficult task.
However, different methods have been proposed for its solution
\cite{efimov,grossmilana,celenza95,bbkr,erf96}.
In the present work we discuss a new approach which is close to that
suggested in \cite{erf96} where an infrared (IR) cut-off  has been
used for the construction of a quark propagator without poles.
In our approach, the quark propagator is of the usual form (with quasiparticle 
pole) but due to the IR cut-off the pole does not lie within the
integration interval for the quark loops.
This method of taking into account the phenomenon of confinement
is based on the idea of combining the NJL and bag models
\cite{bag}.

\begin{figure}
\begin{center}
\epsfbox{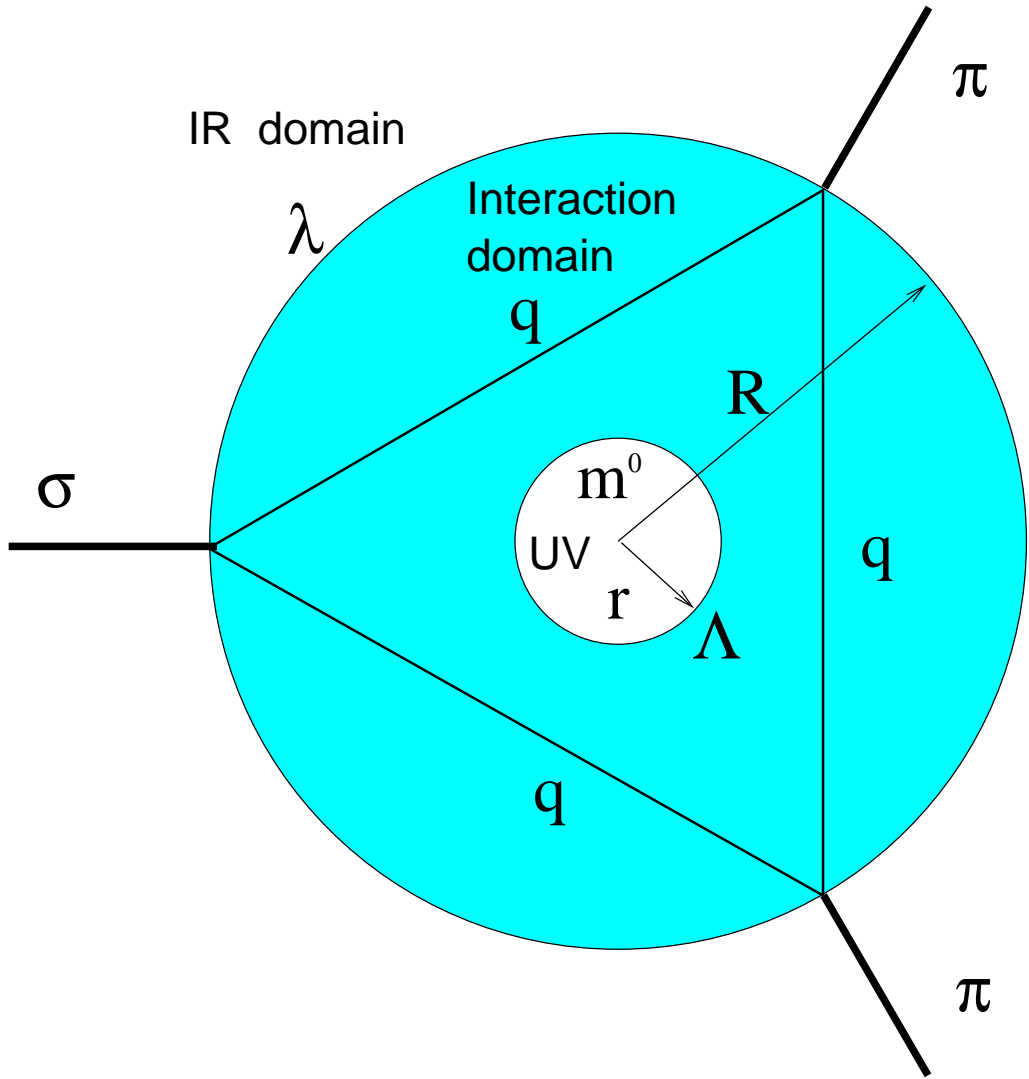}
\end{center}
\caption{Three domains in the momentum space, 
defined by the UV and IR cut-offs.}
\label{domains}
\end{figure}

Thus, together with the ultraviolet (UV) cut-off, which is necessary
for the elimination of the UV divergences, we introduce the IR
cut-off and thereby devide the momentum space into
three domains. In Fig.~\ref{domains} these domains are represented
in the coordinate space.

The first domain corresponds to short distances (large momenta),
where quarks are not confined and the chiral symmetry is not
spontaneously broken. This domain is excluded by the UV cut-off $\Lambda$.

The second domain corresponds to long distances (IR region)
and here we have the confinement of quarks.
We truncate this region from the integration over
the internal momenta in quark loops, following thereby the idea of the bag 
model. For this purpose we introduce a new parameter $\lambda$.

Finally, there remains only the third domain
 $(\lambda^2\leq\vec{\bf p}^2 \leq \Lambda^2)$ where SBCS
takes place, the quark condensate exists and the quark loops have no 
imaginary parts. In other words, quark-antiquark
thresholds do not appear when calculating quark loops
even if the mass of the decaying meson exceeds the effective mass of the 
free quark-antiquark state. Therefore we can use quark propagators with 
constant, momentum  independent, masses (constituent masses).

The first attempt  to construct a NJL model of this type has been made
in \cite{volk98} where  only the scalar and pseudoscalar
mesons were considered.
Now we suggest a more general version  of this model
where the scalar, pseudoscalar and vector mesons can be described and the
possibility of $\pi-a_{1}$-transitions is taken into account.

The paper is organized as follows. In  Sec.~2 we give the effective chiral
quark Lagrangian and the gap equation describing SBCS.
The pion mass formula is also obtained and it is shown that the pion
is a Goldstone boson in the chiral limit.
In Sec.~3 the scalar meson ($\sigma$) is considered,
and it is demonstrated that
the quark loop with two $\sigma$-meson legs does not have
the imaginary part if we
use the IR cut-off. The model parameters are fitted
in Sec.~4. There we consider the decay $\rho \to 2\pi$ through the
triangle quark loop where the
quark-antiquark threshold does not
appear when the IR cut-off is applied.
In Sec.~5 the $\sigma$-meson mass and the decay
$\sigma\to2\pi$ are estimated. 
In the last section we discuss the obtained results and give a short plan
of applying this model for the investigation of the behavior
of the mesons in a hot and dense medium in the vicinity of critical
point. The values of the model parameters, the $\sigma$-meson mass and the 
$\sigma\to 2\pi$ decay width  are given in Table I for different values of 
$\lambda$.

~\section{SU(2)$\times$SU(2) Lagrangian, gap equation and pion mass
formula}

Let us consider an SU(2)$\times$SU(2) NJL model defined by the Lagrangian
\begin{eqnarray}
{\cal L}_q&=& {\bar q}(i{\partial\hspace{-.5em}/\hspace{.15em}}-m^0)q
+ {G_1\over 2} \left[({\bar q} q)^2 +({\bar q}i{\gamma}_5
\vec{\tau} q)^2\right]\nonumber\\
&&-\frac{G_2}{2}\left[ (\bar q\gamma_\mu\vec\tau q)^2+
(\bar q\gamma_5\gamma_\mu\vec\tau q)^2\right]
\label{Lq}~.
\end{eqnarray}
After the bosonization of the four-fermion model (\ref{Lq})
one obtains its equivalent representation in terms of the scalar ($\sigma$),
 pseudoscalar ($\vec{\pi}$), vector ($\rho_\mu$) and
axial-vector ($a_{1\;\mu}$) mesons
\begin{eqnarray}
\label{lmes}
{\cal L}_{meson} &=& -\frac{\tilde{\sigma}^2+\vec{\pi}^2}{2 G_1}
+\frac{\mathop{\vec \rho_\mu }^2+\mathop{\vec{ a}}_{1\;\mu}^2}{2 G_2}
\nonumber\\
&&- i {\rm Tr} \ln\left\{1+\frac{1}{i\partial\hspace{-.5em}/\hspace{.15em} -m}
[\sigma + i \gamma_5\vec{\tau}\vec{\pi}+
\vec\tau{\hat{\vec \rho}}_\mu+\gamma_5\vec\tau{\hat{\vec a}}_{1\;\mu}
]\right\}~.  
\label{lag2}
\end{eqnarray}
Here, the scalar fields $\sigma$ and $\tilde{\sigma}$ are connected by the
relation
\begin{eqnarray}
-m^0 + \tilde{\sigma} &=& -m + \sigma~~,
\end{eqnarray}
where $m^0$ is the current quark mass, $m$ is the constituent quark
mass; and
the vacuum expectation of $\sigma$ vanishes: $\langle \sigma \rangle_0= 0$.
Then, from the condition
\begin{eqnarray}
\frac{\delta {\cal L}}{\delta \sigma}\Bigg|_{\sigma=0, \vec{\pi}=0} =0~~,
\end{eqnarray}
one obtains the gap equation\footnote{
Here, the dependence of the integral $I_1(m)$ on $\lambda$
can be neglected since: 1) the value of the integral is defined
by the UV cut-off $\Lambda$ and 2) $I_1(m)$ does not depend on
external momenta and, therefore, has not the imaginary part.
Hence, there is no need for an IR cut-off.
}
\begin{eqnarray}
m^0&=&m(1-8 G_1 I_1^{(\lambda\,\Lambda)}(m))\nonumber\\
&\approx&m(1-8 G_1 I_1^{(0\,\Lambda)}(m))\nonumber\\
&=&m+2G_1\langle\bar qq\rangle_0  \label{gap}
\end{eqnarray}
where $\langle \bar q q \rangle_0$ is the quark condensate.
$I_1^{(a\,b)}(m)$ is obtained from the
$\Lambda^2$-divergent integral
\begin{eqnarray}
\label{tadpole}
I_1(m)=-i\frac{N_c}{(2\pi)^4}\int
\frac{d^4 k}{m^2 - k^2 -i\varepsilon}
\end{eqnarray}
by applying the 3-dimensional UV ($b=\Lambda$)
and IR ($a=\lambda$) cut-offs
\begin{eqnarray}
I_1^{(\lambda\,\Lambda)}(m)&=&\frac{N_c}{(2\pi)^2}\int^\Lambda_\lambda dk
\frac{k^2}{E(k)}\nonumber\\
&=&\frac{N_c m^2}{8 \pi^2}\left[x\sqrt{x^2+1}-{\rm ln}(x+\sqrt{x^2+1})\right]
\bigg|_{\lambda/m}^{\Lambda/m}~,
\end{eqnarray}
where $E(k)=\sqrt{k^2+m^2}$ and $N_c$ is
the number of colors.

\begin{figure}
\begin{center}
\epsfbox{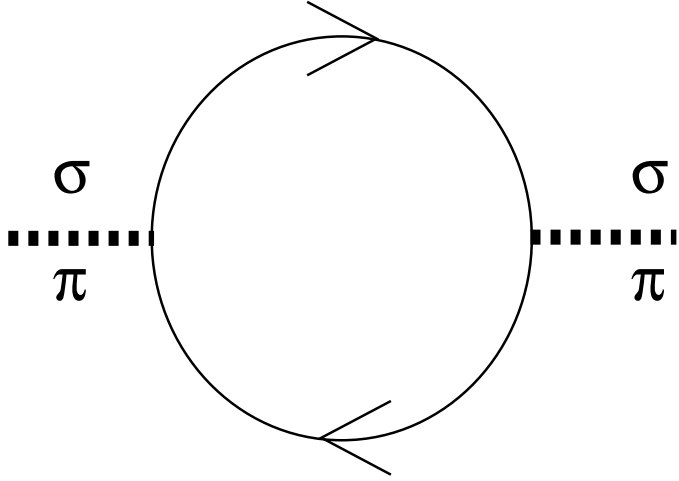}
\end{center}
\caption{The quark-loop diagram for the polarization 
operator of $\sigma$ and $\pi$.}
\label{qloop}
\end{figure}

Now let us consider the free part of the Lagrangian (\ref{lmes}) for pion 
fields in the quark one-loop approximation (see Fig.~\ref{qloop})\footnote{The 
expression in the brackets can be written in the form $1/G_1+\Pi_\pi(p)$, 
where $\Pi_\pi(p)$ is the polarization operator of the pion.}
\begin{eqnarray}
{\cal L}^{(2)}_\pi=-\frac{\vec{\pi}^2}{2}\left\{\frac{1}{
G_1}-8I_1^{(\lambda\,\Lambda)}(m)
-4 p^2 I_2^{(\lambda\,\Lambda)}(p^2,m)\right\}~,
\label{Lpi2}
\end{eqnarray}
where $I_2^{(\lambda\,\Lambda)}(p^2,m)$ is a logarithmically
divergent integral
\begin{eqnarray}
I_2^{(\lambda\,\Lambda)}(p^2,m)
&=&-i\frac{N_c}{(2\pi)^4}\int\frac{d^4 k}{(m^2-k^2-i\varepsilon)
(m^2-(k-p)^2 -i\varepsilon))}\nonumber\\
&=&\frac{N_c}{2 \pi^2}\int_\lambda^\Lambda
dk\frac{k^2}{E(4E^2-p^2+i\varepsilon)}~.
\label{i2}
\end{eqnarray}

In order to express (\ref{Lpi2}) through physical fields,
we renormalize the pions
\begin{eqnarray}
\vec{\pi}= g_\pi(M_\pi)\vec{\pi}^r, \qquad
~g_\pi(M_\pi)\approx g_\pi(0)=[4I_2^{(\lambda\,\Lambda)}(0,m)]^{-1/2}.
\end{eqnarray}
Here, as the pion mass $ M_\pi$ is small, we can approximate the loop 
integral by
$I_2^{(\lambda\,\Lambda)}(p^2=M_\pi^2,m)
\approx I_2^{(\lambda\,\Lambda)}(0,m)$, 
where
\begin{eqnarray}
I_2^{(\lambda\,\Lambda)}(0,m)=\frac{N_c}{8\pi^2}\left[{\rm
ln}(x+\sqrt{x^2+1})-(1+1/x^2)^{-1/2}\right]
\Bigg|_{\lambda/m}^{\Lambda/m}~.
\end{eqnarray}
Moreover, for pions, an additional renormalization
factor $\sqrt{Z}$ appears when we take
 into account the $\pi-a_1$-transitions
\cite{volk86}:
\begin{equation}
\bar g_{\pi}=g_{\pi}\sqrt{Z},\qquad
Z^{-1}=1-\frac{6m^2}{M_{a_1}^2}
\end{equation}
where  $M_{a_1}=1230$ MeV is the mass of the $a_1$-meson.
Thus, we obtain the following expression for the pion mass:
\begin{equation}
M_\pi^2=\bar g_\pi^2\left[\frac{1}{ G_1}-8
I_1^{(\lambda\,\Lambda)}(m)\right]~,
\end{equation}
which can be given the form of the Gell-Mann--Oakes--Renner relation
\begin{equation}
M_\pi^2 \approx -2\frac{m^0\langle\bar qq\rangle_0}{F_\pi^2},
\label{mpi}
\end{equation}
where the Goldberger-Treiman relation (\ref{gold}) and the gap equation 
(\ref{gap}) have been used.
We can see that this pion mass formula is in accordance with the Goldstone
theorem since for $m^0=0$ the pion mass  vanishes  and  it
becomes a Goldstone boson.

\section{The $\sigma$-meson and IR confinement}

The free part of the Lagangian (\ref{lag2}) for the $\sigma$-meson
in the one-loop approximation (see Fig.~\ref{qloop}) has the following form
\begin{eqnarray}
{\cal L}^{(2)}_\sigma=-\frac{{\sigma}^2}{2}\left\{\frac{1}{
G_1}-8I_1^{(\lambda\,\Lambda)}(m)
-4 (p^2-4m^2) I_2(p^2,m)\right\}~.
\label{Lsig2}
\end{eqnarray}
After the renormalization of the $\sigma$ field
\begin{equation}
\sigma=g_\sigma(M_\sigma)\sigma^r,
~g_\sigma(M_\sigma) = [4I_2^{(\lambda\,\Lambda)}(M_\sigma,m)]^{-1/2}~
\end{equation}
we obtain the expression for the $\sigma$-meson mass
\begin{eqnarray}
M_\sigma^2&=& g_\sigma^2(M_\sigma)\left[\frac{1}{ G_1}-8
I_1^{(\lambda\,\Lambda)}(m)\right]+4m^2\nonumber\\
&=&r^2 M_\pi^2+4m^2,\quad \left(r=
\frac{g_\sigma(M_\sigma)}{g_\pi(M_\pi)}\right).
\label{msigma}
\end{eqnarray}
Now let us consider more carefully the integral
$I_2^{(\lambda\,\Lambda)}(M_\sigma,m)$:
\begin{eqnarray}
I_2^{(\lambda\,\Lambda)}(M_\sigma^2,m)=
\frac{N_c}{2\pi^2} \int_{\lambda}^{\Lambda}
\! d k \frac{k^2}{E(4E^2-M_\sigma^2+i\varepsilon)}
\label{i2sigma}
\end{eqnarray}
When $\lambda=0$ this integral has an imaginary part.
Indeed, the  integrand in (\ref{i2sigma})
is singular when its denominator
is equal to zero:
\begin{eqnarray}
4E^2-M_\sigma^2=4k^2-r^2M_\pi^2=0. \label{id}
\end{eqnarray}
The imaginary part appears if the singularity ($k=\frac{r}{2}M_\pi$)
lies within the integration interval.
Therefore, if we apply the IR cut-off
\begin{eqnarray}
\lambda=c\,m\quad{\rm where}\quad
c>\frac{r M_{\pi}}{2m}, \label{lambdacond}
\end{eqnarray}
then $\Lambda>k>\lambda>\frac{r}{2}M_\pi$ and
the integral is real,
which means the absence of the quark-antiquark threshold, or
confinement.

\section{Model parameters}

In this model we have five parameters: the constituent quark mass $m$, 
the scalar (pseudoscalar) four-quark coupling constant $G_1$,
the vector (axial-vector) four-quark coupling constant $G_2$,
the 3-momentum UV cut-off parameter $\Lambda$ and 
the 3-momentum IR cut-off parameter $\lambda$.

Since we will use for the determination of these parameters only the four 
pion and $\rho$ meson observables \cite{PartProp}: 
$M_\pi=140$ MeV, $F_\pi=92.4$ MeV, $M_\rho=770$ MeV and $g_\rho^{\rm exp}=6.14$,
the IR cut-off $\lambda$ is an arbitrary parameter of our model
which is to satisfy the condition (\ref{lambdacond}). 
We have checked the sensitivity of the model to a variation of $\lambda$ and 
the results are summarized in Table I. 
We will consider $\lambda=m$ as optimal choice since it does not only give 
acceptable $\sigma$ meson properties (see next Section). 
\footnote{This is also 
suggested from a comparison with the simple confining model by Munczek and 
Nemirovsky \cite{mn83}, where in the chiral limit the lower limit of 
quark four-momenta $p$ for which quarks are deconfined is given by the 
dynamical quark mass function at $p^2=0$.}
 
In order to fix the output parameters we use the following four
equations:
\begin{enumerate}
\item[1)] The Goldberger--Treiman relation
\begin{equation}
\frac{m}{F_{\pi}}=\bar g_{\pi}(0)=g_\pi(0)\sqrt{Z}, \label{gold}
\end{equation}
\item[2)] The decay width of the process $\rho \to 2\pi$. The amplitude
of this process is of the form
\begin{equation}
T_{\rho\to2\pi}=i\frac{g_{\rho}^{\rm exp}}{2}
(p_{\pi^+}-p_{\pi^-})^\nu \rho^0_{\nu}\pi^+\pi^-.
\end{equation}

In the one-loop approximation (see Fig.~\ref{rs2pi}), we obtain the
following expression for $g_{\rho}^{\rm exp}$
\begin{eqnarray}
g_{\rho}^{\rm exp}&=&
Z^{-1}g_{\rho}(M_{\rho})\bar g_{\pi}^2(M_{\pi})
\left[4I_2^{(\lambda\;\Lambda)}(0,m)+\Delta \right]\\
&=&\left(\frac23 I_2^{(\lambda\,\Lambda)}(M_{\rho}^2,m)\right)^{-1/2}
\left[1+\frac{\Delta}{4I_2^{(\lambda\,\Lambda)}(0,m)}\right]~,
\end{eqnarray}
where $g_{\rho}(M_\rho)=[\frac23
I_2^{(\lambda\,\Lambda)}(M_{\rho}^2,m)]^{-1/2}$
(see \cite{volk86}) and $\Delta=\frac{3}{8\pi^2
}\left(1+\frac{2M_\pi^2}{3m^2}\right) $ is the finite part
 of the quark triangle diagram (Fig.~\ref{rs2pi}).
The factor $Z^{-1}$ appears due to the
$\pi-a_1$-transitions (see \cite{volk86}).
From these two equations one can find $m$ and $\Lambda$.
\item[3)] The coupling constant $G_1$ is defined by the
mass formula
\begin{eqnarray}
M_{\pi}^2=
\bar g_{\pi}^2\left[{1\over G_1}-8 I_1^{(\lambda\;\Lambda)}(m)\right]~.
\end{eqnarray}
\item[4)] The coupling constant $G_2$  is found from  the  mass
formula for $M_\rho$ \cite{volk86}
\begin{eqnarray}
M_{\rho}^2=\frac{g^2_{\rho}(M_\rho)}{G_2}=
\frac{3}{8G_2 I_2^{(\lambda\,\Lambda)}(M_\rho^2,m)}.
\end{eqnarray}
\end{enumerate}
By means of the gap equation (\ref{gap}) we define the current quark mass
$m^0$. The result of the above described parameter fixing procedure is 
summarized in Table I for different choices of the IR cut-off 
$\lambda$ given by the ratio $\lambda/m$.
\begin{table}
\caption{Model parameters, the mass of the $\sigma$-meson and
its decay width for different values of the ratio of the IR
cut-off to the constituent quark mass $\lambda/m$.}
\begin{indented}
\item[]\begin{tabular}{@{}ccccccccc}
\br
$\lambda/m$ & m & $m^0$ &$\Lambda$  & $G_1$ & $G_2$ & $-\langle \bar 
qq\rangle^{1/3}$ & $M_\sigma$ & $\Gamma_\sigma$ \\
 & [MeV] & [MeV] & [GeV] & [GeV]$^{-2}$ & [GeV]$^{-2}$ & [MeV] & [MeV] & [MeV] 
\\
\mr
0.7 & 330 & 1.54 & 1.11 & 3.108 & 9.683 & 298 & 670 & 1280 \\
0.8 & 310 & 1.58 & 1.13 & 2.967 & 8.629 & 296 & 630 & 970 \\
0.9 & 300 & 1.49 & 1.18 & 2.698 & 8.949 & 302 & 608 & 830 \\
1.0 & 290 & 1.45 & 1.21 & 2.555 & 9.222 & 304 & 590 & 721 \\
1.1 & 280 & 1.37 & 1.26 & 2.339 & 9.322 & 309 & 570 & 614 \\
1.2 & 270 & 1.32 & 1.33 & 2.079 & 9.273 & 318 & 550 & 536 \\
\br
\end{tabular}
\end{indented}
\label{tab1}
\end{table}

\section{The $\sigma$-meson mass and the decay $\sigma \to 2 \pi$}

\begin{figure}[b]
\begin{center}
\epsfbox{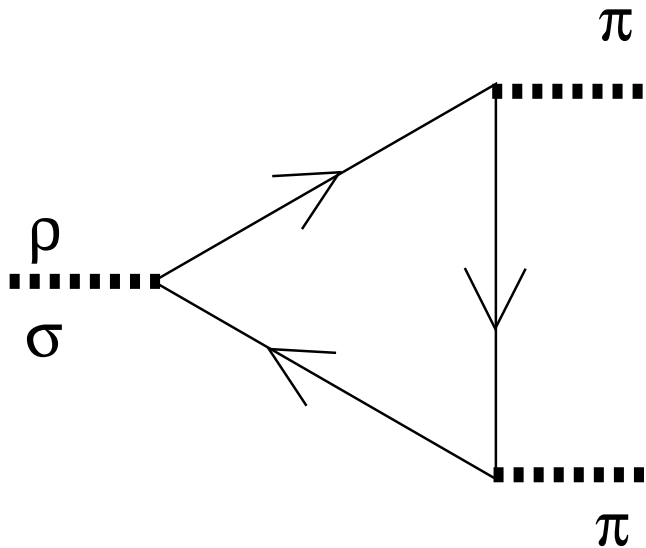}
\end{center}
\caption{The triangle quark diagram describing the decay of the 
$\rho$- and $\sigma$-mesons into two pions.}
\label{rs2pi}
\end{figure}

The mass of the $\sigma$-meson is given by  eq. (\ref{msigma}).
Using this formula for the IR cut-off $\lambda=m$ we obtain
\begin{equation}
M_\sigma=590\;{\rm MeV}.
\label{msig2}
\end{equation}
The decay $\sigma \to 2 \pi$ occurs through the quark triangle
diagram (see Fig.~\ref{rs2pi}).

This diagram also satisfies the confinement condition (\ref{lambdacond})
if we use the IR cut-off
$\lambda= m$.
The amplitude of the process  $\sigma \to 2 \pi$ has the form
\begin{eqnarray}
T_{\sigma\to2\pi}&=&8 m g_\sigma(M_\sigma)\bar g_\pi^2(M_\pi)
[I_2^{(\lambda\;\Lambda)}(M_\sigma^2,m)+
{\cal J}(M_\sigma,M_\pi,m)] \sigma \vec{\pi}^2~~,
\end{eqnarray}
where
\begin{eqnarray}
{\cal J}(M_\sigma,M_\pi,m)&=& \left.\frac{1}{2}(M_\sigma^2-2M_\pi^2)
I_3(p_1,p_2,m)\right|_{p_1^2=p_2^2=M_\pi^2}
\end{eqnarray}
and
\begin{eqnarray}
I_3(p_1,p_2,m)&=&\frac{-i N_c}{(2\pi)^4}\int
\frac{d^4k}{(k^2-m^2)((k+p_1)^2-m^2)((k-p_2)^2-m^2)}.
\end{eqnarray}
Neglecting the external pion momenta in $I_3(p_1,p_2,m)$ we obtain
\begin{eqnarray}
I_3(p_1,p_2,m)
&\approx&\frac{-i N_c}{(2\pi)^4}\int \frac{d^4k}{(k^2-m^2)^3}
=-\frac{3 N_c}{32\pi^2}\int_\lambda^\Lambda dk \frac{k^2}{E^5}\nonumber\\
&=&-\frac{N_c}{32\pi^2m^2}\left[\left(1+\frac{m^2}{\Lambda^2}\right)^{-3/2}-
\left(1+\frac{m^2}{\lambda^2}\right)^{-3/2}\right]~.
\end{eqnarray}
Then the decay width of  the  $\sigma$-meson  is
equal to
\begin{eqnarray}
\Gamma_{\sigma\to2\pi}&=&\frac{3}{2\pi}
\left(\frac{m^3}{g_\sigma(M_\sigma) F_\pi^2}\right)^2
\frac{(1+\delta)^2}{M_\sigma^2}\sqrt{M_\sigma^2-4M_\pi^2}
\approx 720~ {\rm MeV}~,
\label{sigwidth}
\end{eqnarray}
where
\begin{equation}
\delta=\frac{{\cal J}(M_\sigma,M_\pi,m)}{I_2^{(\lambda\;\Lambda)}(0,m)}
\approx -0.22 .
\end{equation}
Therefore,  one can see that our estimates for the $\sigma$-meson  mass  
and its decay  width  are  in  agreement  with  the  experimental data
\cite{PartProp} (see also \cite{Ishi_96,Svec_92}):
\begin{equation}  
M_\sigma^{\rm exp} = (400 - 1200)~ {\rm MeV}~,~~ 
\Gamma_\sigma^{\rm  exp} = (600 - 1000)~ {\rm MeV}~.
\end{equation}
From this one can conclude that the NJL model with the IR cut-off satisfies 
both of the low energy theorems together with SBCS and describes the 
low-energy physics of the scalar, pseudoscalar and vector mesons.

\section{Discussion and conclusion}

In this paper we have investigated the extension of the NJL model for the
light nonstrange meson sector of QCD, where the interaction of
$u$- and $d$-quarks is represented by four-fermion vertices and the
phenomenon of quark confinement is taken into account. This extension
of the NJL model  describes the properties of the $\pi$, $\rho$ and
$\sigma$-mesons in good agreement with the experiment and with the low-energy 
theorems.
The model parameters are obtained by fitting the model so that it reproduces 
the experimental values of the pion and $\rho$-meson masses, the pion decay 
constant $F_\pi$ and the $\rho$-meson decay constant $g_\rho$.
Moreover, it was shown that for the $\pi$-, $\rho$- and $\sigma$-mesons the 
non-physical quark-antiquark thresholds do not appear if the IR cut-off is 
applied.

The prediction of the $\sigma$-meson mass and its decay width for different 
values of IR cut-off is given in Table I together with the model parameters. 
From this table and the experimental data
\cite{PartProp} one can conclude that the parameter $\lambda$ is
allowed to have values within the interval $0.8 \leq \lambda/m \leq 1.2$.
The value of the current quark mass turned out to be too low because
in our model the quark condensate is greater than its conventional
value $-(250 {\rm MeV})^3$. This can be seen from eq.~(\ref{mpi}).

Insofar as the NJL model is a semiphenomenological model based on
the effective chiral four-quark interacion motivated by QCD on
the phenomenological level, the introduction of quark confinement
in our model by means of an IR cut-off (without considering the
gluon exchanges, instanton interactions etc.) is in the spirit of
this model.

An interesting  application
of our model is the description of meson properties in a hot
and dense medium. The standard NJL model has been already used
for this purpose \cite{vogl,hatsuda,munchov}, where the temperature
dependence of the masses of quarks and mesons and of the Yukawa coupling
constants was found.
The IR cut-off $\lambda$ is expressed through the constituent quark
mass $m$ (or the quark condensate $\langle \bar qq\rangle_0$), which
decreases when the temperature $(T)$ and chemical potential $(\mu)$
increase.  Therefore, the IR
cut-off will also decrease with $T$ and $\mu$.
This will at length result in the deconfinement of quarks
near the critical point. The temperature at which
deconfinement takes place
can be found from the condition (see also eq.~(\ref{id}))
\begin{equation}
4E^2_{\rm low}-M^2_{\rm meson}=4(1+c^2)m^2-M^2_{\rm meson}\leq 0
\label{deconfcond}
\end{equation}
where $E_{\rm low}$ is the lowest energy of the quark in the quark loop.
Then, for the pion, the temperature of deconfinement follows from the
condition (\ref{deconfcond})
\begin{equation}
m(T^{\rm dec}_{\pi})\approx \frac{M_\pi}{2\sqrt{1+c^2}} ~.
\end{equation}
In the vicinity of the critical point, the constituent quark mass decreases 
with $T$ very quickly, and the pion mass slowly increases 
(see \cite{munchov}). Therefore, when the constituent quark mass is as light 
as 40--50 MeV, the decay of pion into free quarks becomes possible.

For the $\sigma$-meson we obtain a lower value of $T^{\rm dec}$:
\begin{equation}
m(T^{\rm dec}_{\sigma})\approx \frac{M_\pi}{c}r.
\end{equation}

The decay channel $\sigma\to 2\pi$ is closed  when
$M_{\sigma} \leq 2M_{\pi}$
(see eq.~(\ref{sigwidth})) and then
\begin{equation}
m(T_{\sigma\not\to2\pi})\approx\frac{\sqrt{3}}{2}M_{\pi}r\sim 100 \; {\rm MeV}.
\end{equation}

Note that, first of all, the quark deconfinement occurs for the
$\rho$-meson, where we have the lowest $T^{\rm dec}$ (see
Eqs.~(\ref{deconfcond})
and \cite{munchov})
\begin{equation}
m(T^{\rm dec}_{\rho})\approx \frac{M_{\rho}}{2\sqrt{1+c^2}}
\approx 250\; {\rm MeV}.
\end{equation}

Thus, we have the following picture. At low $T$ and $\mu$ the $\sigma$-meson
is unstable since it has a large decay width (\ref{sigwidth}) into two pions.
The pion is stable since electroweak decay channels can be neglected here in 
comparison to the strong ones. 
The $\rho$-meson takes an intermediate position between 
$\sigma$ and $\pi$, having the $\rho\to2\pi$ strong decay width 150 MeV.
With $T$ increasing, there opens an additional decay channel: 
$\rho\to\bar qq$.   
At higher $T$, there exists an interval in the temperature scale where the 
decays of $\sigma$ both into $2\pi$ and into $\bar{q} q$ are forbidden, and 
the $\sigma$-meson turns out to be a stable particle.
Next, when $T > T^{\rm dec}_{\sigma}$ the $\sigma$-meson is allowed
to decay into a $\bar{q} q$ pair and only the pion remains to be stable.
Finally, near the critical point, when $T \geq T^{\rm dec}_{\pi}$, all the
particles decay into free quarks.

The whole process of deconfinement is reflected in the following sequence of 
inequalities
\begin{equation}
T^{\rm dec}_{\rho}<T_{\sigma\not\to2\pi}<T_{\sigma}^{\rm dec}<T^{\rm
dec}_{\pi}
\end{equation}
where we have four different temperatures separating different
phases.
Therefore, one can see that the transition of the hadron matter to
the quark-gluon plasma occurs not abruptly but in a smooth manner.

In our further work we are going to make a more careful investigation of 
these processes, which can play an important r\^ole for the explanation and 
prediction of signals coming from ultrarelativistic heavy-ion collisions 
and witnessing the chiral symmetry restoration and the quark deconfinement in 
hadron matter at the transition to the quark-gluon plasma.
Of particular interest is the study of quark substructure effects on
$\rho$ meson properties which is possible within the present model since 
their modifications at the suspected QCD phase 
transition are cruical for understanding, e.g., the low-mass 
dilepton enhancement as observed by the CERES collaboration \cite{CERES}.

\section*{ Acknowledgments}

This work has been supported by RFFI Grant N 98-02-16135 and the 
Heisenberg-Landau program, 1998--1999. GB and MKV acknowledge support 
by the Max-Planck-Gesellschaft.

\end{document}